\documentclass{PoS}

\usepackage[applemac]{inputenc}
\usepackage{caption}
\usepackage{subcaption}
\usepackage{lineno}

\title{Full-Sky Analysis of Cosmic-Ray Anisotropy with IceCube and HAWC}

\ShortTitle{Full-Sky Analysis of Cosmic-Ray Anisotropy with IceCube and HAWC}

\author{The HAWC Collaboration$^{1}$, The IceCube Collaboration$^{2}$   
\\
{\it 
$^1$ \href{http://www.hawc-observatory.org/collaboration/icrc2015.php}{www.hawc-observatory.org/collaboration/icrc2015.php} \\
$^2$ \href{http://icecube.wisc.edu/collaboration/authors/icrc15_icecube}{http://icecube.wisc.edu/collaboration/authors/icrc15\_icecube}\\
}

E-mail: \email{juancarlos@icecube.wisc.edu}
}

\abstract{During the past two decades, experiments in both the Northern and Southern hemispheres have observed a small but measurable energy-dependent sidereal anisotropy in the arrival direction distribution of galactic cosmic rays. The relative amplitude of the anisotropy is $10^{-4} - 10^{-3}$. However, each of these individual measurements is restricted by limited sky coverage, and so the pseudo-power spectrum of the anisotropy obtained from any one measurement displays a systematic correlation between different multipole modes $C_\ell$. To address this issue, we present the preliminary status of a joint analysis of the anisotropy on all angular scales using cosmic-ray data from the IceCube Neutrino Observatory located at the South Pole ($90^\circ$ S) and the High-Altitude Water Cherenkov (HAWC) Observatory located at Sierra Negra, Mexico ($19^\circ$ N). 
We describe the methods used to combine the IceCube and HAWC data, address the individual detector systematics and study the region of overlapping field of view between the two observatories.

\vspace{4mm}
{\bf Corresponding authors:}
\speaker{Juan Carlos D\'iaz-V\'elez$^{1a,b}$},
Dan Fiorino$^{2a}$,
Paolo Desiati$^{a}$,
Stefan Westerhoff$^{a}$,
Eduardo de la Fuente$^{b}$\\
{\it
$^1$ juancarlos@icecube.wisc.edu \\
$^2$ dan.fiorino@icecube.wisc.edu \\
} \\
     \llap{$^a$}Wisconsin IceCube Particle Astrophysics Center (WIPAC) and Department of Physics, 
     University of Wisconsin--Madison, Madison, WI 53706, USA \\
     \llap{$^b$}Centro Universitario de Ciencias Exactas e Ingenier\'ias, y Centro Universitario de los Valles, Universidad de Guadalajara, Guadalajara, Jalisco 44130, M\'exico
}

\FullConference{The 34th International Cosmic Ray Conference,\\
		30 July- 6 August, 2015\\
		The Hague, The Netherlands}

\begin{document}

\section{Introduction}
Over the last few decades, several studies have measured
appreciable variation in the intensity of cosmic rays of
medium and high energies as a function of right ascension.
An anisotropy with an amplitude of  $10^{-4}$ was first observed at energies of order 1 TeV by a number 
of experiments including the Tibet AS$\gamma$ \cite{Tibet:2005jun}, Super-Kamiokande \cite{SuperK:2007mar}, 
Milagro \cite{Milagro:2008nov}, EAS-TOP \cite{Aglietta:2009feb}, MINOS \cite{MINOS:2011icrc}, ARGO-YBJ \cite{ARGO:2013jun}, 
and HAWC \cite{HAWC:2014dec} experiments in the Northern Hemisphere
and IceCube \cite{IceCube:2010aug,IceCube:2011oct,IceCube:2012feb} and its surface air shower array
IceTop \cite{IceCube:2013mar} in the Southern Hemisphere.
In both hemispheres, the observed anisotropy has two main features: a large-scale structure with an amplitude of about 
$10^{-3}$,  and a small-scale structure with an amplitude of $10^{-4}$ with a few 
localized regions of cosmic-ray excesses and deficits of angular size $10^\circ$ to $30^\circ$. 

The origin of this anisotropy is not yet well understood since it
is expected that cosmic rays should lose any correlation with their original direction due to
diffusion as they traverse through interstellar magnetic fields.
There are several theories regarding the possible origin of this
anisotropy including ones that postulate a heliospheric origin
\cite{Lazarian:2010sq,bib:desiati2013} though it is also possible that
the origin of anisotropy is due to characteristics
of the interstellar magnetic field at distances less than 1 parsec or
even the diffuse flow of nearby galactic sources \cite{Desiati:CRA13}. 
Others have proposed scenarios where the anisotropy results from the distribution of cosmic ray sources in the Galaxy and of their diffusive propagation
\cite{Erlykin:2006apr, Blasi:2012jan, Ptuskin:2012dec, Pohl:2013mar, Sveshnikova:2013dec, Kumar:2014apr, Mertsch:2015jan}.

IceCube and HAWC data can be combined at the same energy to study the full-sky anisotropy.
Important information can be obtained on the power spectrum at low-$\ell$ (large scale), which is the region most affected by partial sky-coverage of one experiment only.

\section{The Dataset}

IceCube is a km$^3$ neutrino detector located at the geographic South Pole. It is composed of 5160
Digital Optical Modules (DOMs) deployed at depths between 1450 m and 2450 m below the surface of the ice sheet.
It detects the Cherenkov radiation emitted by relativistic charged particles as they propagate through the ice.
The event rate varies between 2 kHz and 2.4 kHz, with the modulation caused by 
seasonal variations of the stratospheric temperature. 
The detected muon events are generated by primary cosmic-ray particles with 
median energy of 20 TeV, as determined by simulations. 
The estimated median angular resolution for this dataset from simulation is $3^\circ$ \cite{Westerhoff:2015aug}.


The HAWC Observatory is a 22,000 m$^2$ dense array of 
250 water Cherenkov detectors (WCDs) with 1200 photomultipliers (PMTs). Each WCD
contains four photomultipliers per
tank: one central high-quantum-efficiency
10-inch PMT and three 8-inch PMTs.  
The trigger rate in HAWC-250 is approximately
16 kHz. With 250 WCDs
the angular resolution of the air shower reconstruction is between
$0.3^\circ$ and $1.5^\circ$. 
Poorly reconstructed events are reduced by requiring at least 6\% of the PMTs are hit.
From simulations we estimate that the 
median energy of the data set is about 2 TeV \cite{Benzvi:2015aug}.

\begin{figure}[h]
\begin{center}
\includegraphics*[width=.75\textwidth]{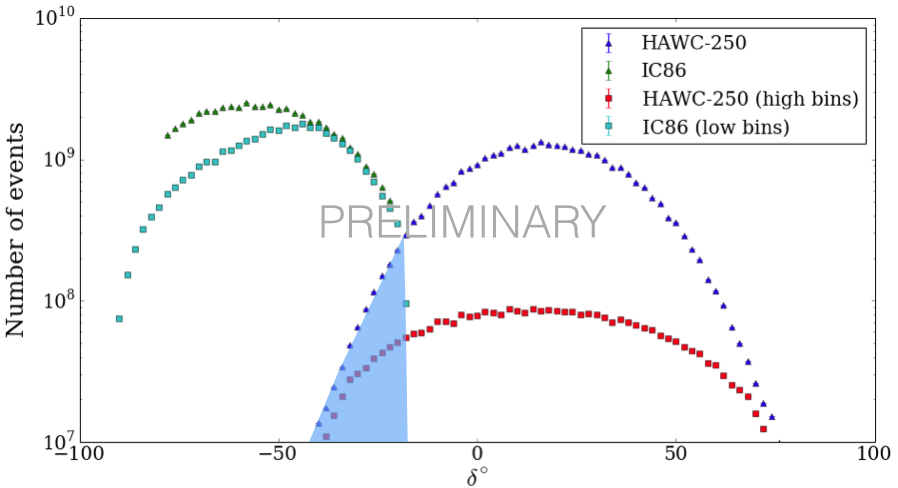} \end{center}
\caption[Detector acceptance]{
Distribution of events as as a function of declination for IceCube and HAWC.
The shaded area corresponds to the overlapping region for both experiments. 
Triangles correspond to the full energy spectrum and squares correspond to the same datasets after applying energy cuts. 
Restricting datasets to overlapping energy bins significantly reduces statistics for HAWC. 
}
\label{fig:detector_acceptance} 
\end{figure}

\begin{figure*}[h]
        \centering     
        \begin{subfigure}[b]{0.47\textwidth}
                \includegraphics[width=\textwidth]{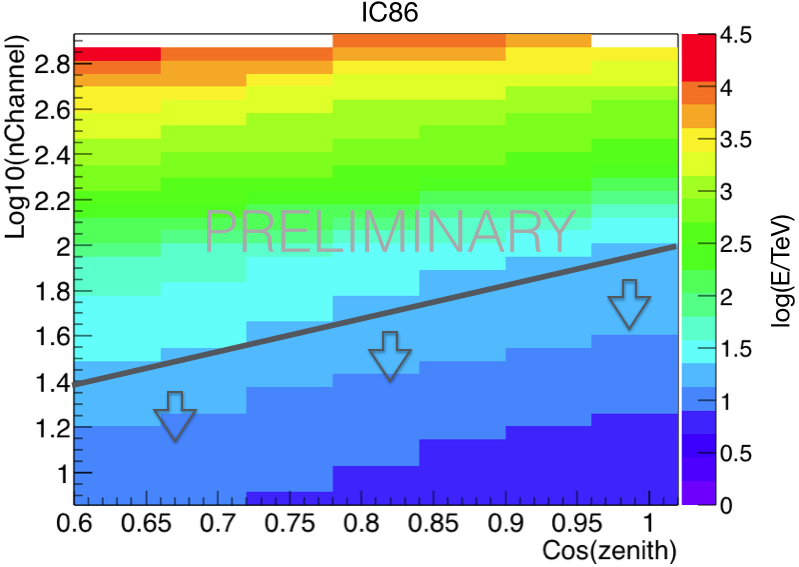}
                \label{fig:nchan_costh:ic}
        \end{subfigure}
        \begin{subfigure}[b]{0.47\textwidth}
                \includegraphics[width=\textwidth]{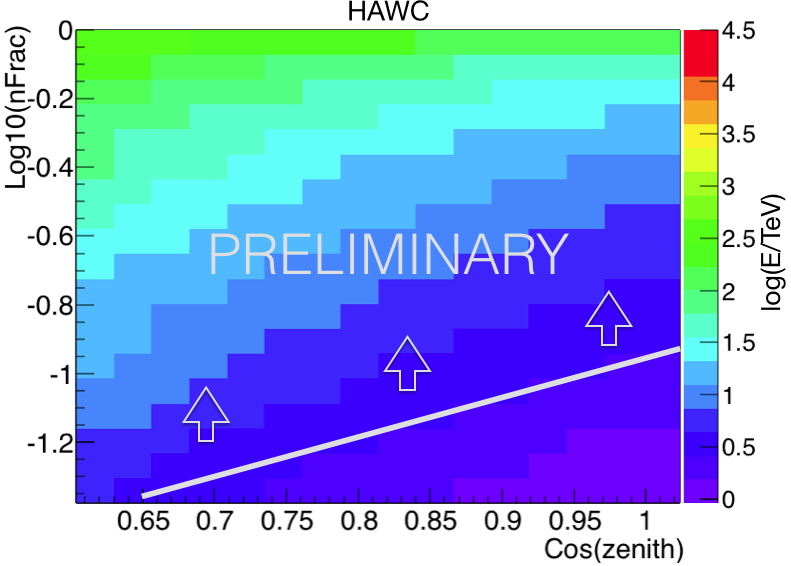}
                \label{fig:nchan_costh:hawc}
        \end{subfigure}
        \caption[Energy cuts]{Energy cuts for IceCube (left) and HAWC (right) datasets. 
        The two figures show the logarithm of the mean energy as a function of cosine zenith angle and number 
        of hit channels (IceCube) or fraction of hits to number of available channels (HAWC) used as an energy proxy. 
        Data are selected such that the two datasets are compatible with a median energy of ~10 TeV.}
        \label{fig:nchan_costh}
\end{figure*}

Data selected for preliminary analysis come from the third year of IceCube in its final configuration of 86 strings (IC86), as well as four months of HAWC in its preliminary configuration of 250 tanks (HAWC-250). Table \ref{tab:observatories} shows the characteristics of both detectors next to each other.  
Only continuous sidereal days of data were chosen for 
these analyses in order to reduce the bias of uneven exposure
along right ascension. 
For the purpose of studying systematics effects, we also use data from the HAWC-111 configuration that was collected between June 2013 and February 2014 by HAWC, when the observatory was operated with 111 WCDs during detector construction. 
The changing detector configuration within the HAWC-111 dataset makes it difficult to get a good energy estimation and is thus not suitable for the overall analysis.
Figure \ref{fig:detector_acceptance} shows the distribution of data as a function of declination. There is a very narrow region of overlap between the two detectors. The statistics in HAWC-250 are  comparable to one year of IC86. As evident from Table \ref{tab:observatories}, there is also a difference in median energy of both experiments. A further complication is that the median energy
grows as a function of zenith angle so that the region of overlap corresponds to the maximum median energy of both detectors. 
Figure \ref{fig:nchan_costh} show this dependency and illustrates the way to select consistent data between the two detectors.

\begin{table}[h]
\centering  
\footnotesize
\begin{tabular}{p{0.23\textwidth}|p {0.33\textwidth} | p{0.30\textwidth}}
{}&{\small \bf{IceCube}} & {\small \bf {HAWC}} \\  \hline
Hemisphere & Southern  & Northern  \\
Latitude& -90$^\circ$ &  19$^\circ $\\
Trigger rate & 2.5 kHz  &  10 kHz  \\
Detection method &  muons produced by CR and neutrinos  &  cascades produced by CR and $\gamma$  \\
Median primary energy & $10$ TeV  &  $2$ TeV  \\
Approx. angular resolution &  $2^\circ - 3^\circ$ &  $0.3^\circ - 1.5^\circ$ \\
Field of view  &  -90$^\circ$/-20$^\circ$, ${\sim}$4 sr (always observes the same sky) &  -30$^\circ$/64$^\circ$,  ${\sim}$2 sr (8 sr observed)\\
Livetime & 362.2 days  ($6.2 \times 10^{10}$ events) &  52 days  ($3.3  \times 10^{10}$ events ) 
\end{tabular}
\caption [Comparison of observatories] {
Comparison of the IceCube and HAWC datasets.
} \label{tab:observatories}
\end{table}

\section{Analysis}
We compute the relative intensity as a function of equatorial
coordinates ($\alpha$,$\delta$) by binning the sky into an
equal-area grid with a resolution of 0.2$^\circ$ per bin using the 
HEALPix library~\cite{Gorski:2005apr}. 
In order to produce residual maps of the anisotropy of the arrival
directions of the cosmic rays, we must have a 
description of the arrival direction distribution if the
cosmic rays arrived isotropically at Earth $\langle N \rangle (\alpha,\delta)_i$. 
We calculate this expected
flux from the data themselves in order to account for rate variations 
in both time and viewing angle.
While the estimation of this reference map is calculated differently in IceCube and HAWC, these two methods are compatible. 
For HAWC, it is produced using the direct integration
technique as described in \cite{Benzvi:2015aug}, and  in the case of IceCube the same is accomplished by time-scrambling events in local coordinates \cite{IceCube:2010aug} within a time window
$\Delta t$.
Once the reference map is obtained, we calculate the
deviations from isotropy by computing the relative intensity 
\begin{equation}\label{eq:relint}
\delta I(\alpha,\delta)_i = \frac{N(\alpha,\delta)_i - \langle N \rangle(\alpha,\delta)_i}{\langle N \rangle (\alpha,\delta)_i}~,
\end{equation}
where $N(\alpha,\delta)_i,  \langle N \rangle(\alpha,\delta)_i$ are the number of observed events and the number of reference events in the $i^{th}$ bin of the map, respectively.
The relative intensity gives the amplitude of deviations from the isotropic
expectation in each angular bin $i$. 

This analysis method can be sensitive to any maximum angular 
scale through the choice of $\Delta t$. Due to the sampling of the
reference map along lines of right ascension, the maximum angular
scale shrinks as $1/\cos(\delta)$. Only a choice of 24 hours ensures
a uniform angular scale as a function of $\delta$. 
To eliminate larger structures, a multipole fit can be subtracted to access lower angular 
scales while preserving the maximum angular scale throughout the map.

Analyses of data from Earth-based experiments with partial sky coverage suffer from systematic effects and statistical uncertainties 
of the calculated angular power spectrum. 
An additional limitation is the fact that the combined map will  suffer from the fact that this analysis is only sensitive to projections 
of large-scale structure onto the right ascension.

\subsection {Systematic Checks}
\begin{figure}[ht]
\begin{center}
\includegraphics*[width=.85\textwidth]{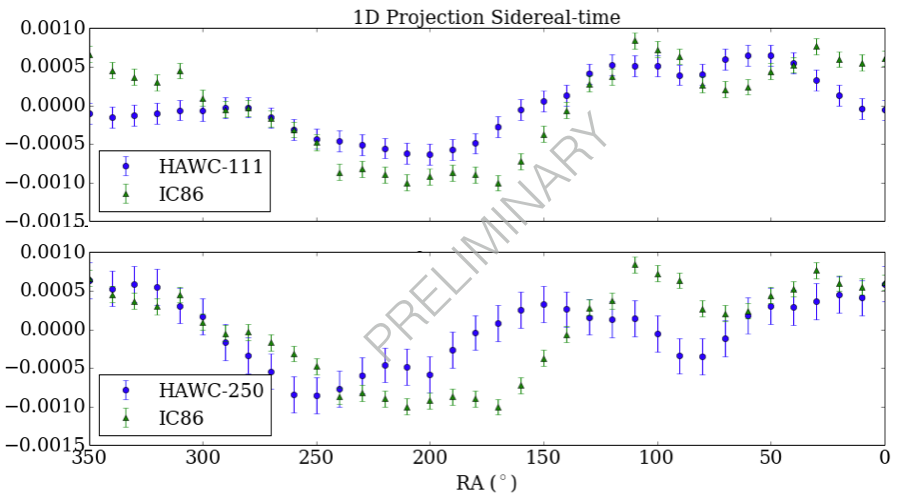} \end{center}
\caption[
1D projection of large-scale anisotropy.]{
One-dimensional RA projection of the relative intensity of cosmic rays within the overlapping $\delta$ region (between -30$^\circ$ and -20$^\circ$)
for HAWC-111 (top) and HAWC-250 (bottom) compared to IC86 data. 
HAWC-111 (which covers a larger portion of a year) shows better agreement with 
IC86 while HAWC-250 shows a significant deviation in phase. 
Better agreement is expected with the accumulation of more statistics though some small-scale differences 
can be expected as a result of contamination from mis-reconstructed events belonging to other declination bands.
} 
        \label{fig:sidereal}
\end{figure}

\begin{figure}[h]
\begin{center}
\includegraphics*[width=.83\textwidth]{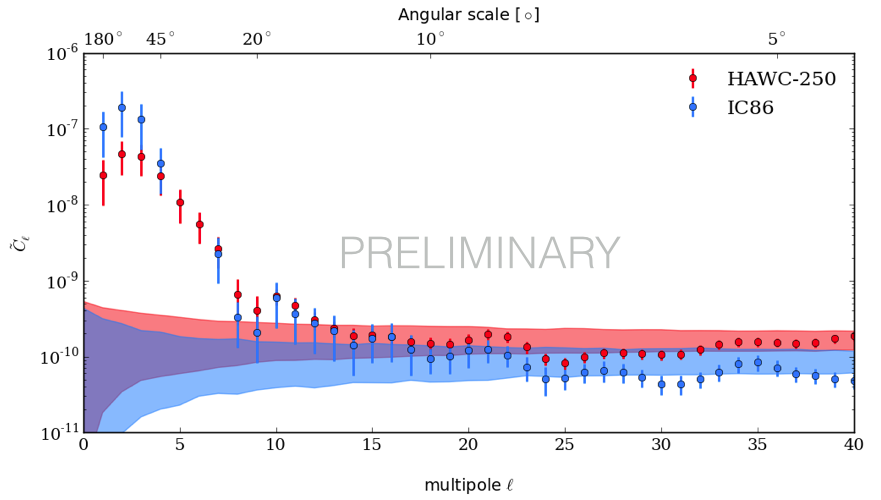} \end{center}
\caption[Power spectrum]{
Power spectrum for IC86-III and HAWC-250 datasets. Solid color bands represent the power spectra for isotropic sky maps at the 90\% confidence level. 
} 
        \label{fig:power-spectrum}
\end{figure}

Sidereal anisotropy can be distorted by a yearly modulation in solar time
unless the data are uniformly distributed over an integer number
of complete years.  One such modulation is the ``solar dipole'',
an observable anisotropy induced by the motion of the Earth through the Solar wind.
The frame referred to as ``anti-sidereal'' time is a non-physical reference system which is obtained by inverting
the sign on the conversion of solar time to sidereal time (adding 4 minutes per solar day) so the anti-sidereal year has 364.25 days. 
This anti-sidereal time frame can be used to study systematic effects caused by seasonal variations \cite{Farley1954}.
We produce a skymap where anti-sidereal time is used instead of sidereal time in the coordinate transformation 
from local detector coordinates to ``equatorial'' coordinates. 
\begin{figure*}[ht]
\begin{center}
\includegraphics*[width=0.85\textwidth]{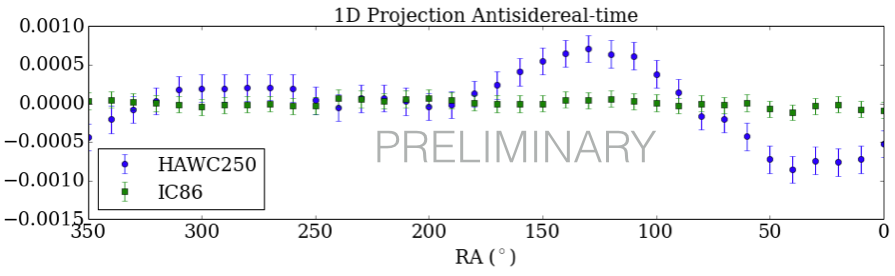} \end{center}
\caption[1D projection of anti-sidereal distributionl]{
One-dimensional projection of the relative intensity of cosmic rays within the overlapping $\delta$ region (between -30$^\circ$ and -20$^\circ$) as a function of 
RA as defined in the anti-sidereal coordinate system, a non-physical reference system that is obtained by inverting the sign 
on the conversion of solar time to sidereal time. 
This frame can be used to study systematic effects caused by seasonal variations. 
The presence of a seasonal effect such as a solar dipole is apparent in the HAWC-250 dataset as a significant deviation from a flat distribution.
}
\label{fig:antisidereal} 
\end{figure*}
Figure \ref{fig:antisidereal} shows the distribution of the relative intensity of CR
under anti-sidereal reference. There is an apparent modulation in the HAWC data due to the
fact that these data do not cover a full year. 
The effect of this modulation can be seen in the large-scale anisotropy when compared to IC86. Figure \ref{fig:sidereal} 
shows the one-dimensional projection in right ascension for HAWC-111 and HAWC-250 compared to IC86. HAWC-111 (which covers a larger portion of a sidereal year) shows better agreement with IC86 while HAWC-250 shows a significant deviation in phase. Better agreement is expected with the accumulation of more statistics though, since the two experiments are observing different portions of the sky, there is no expectation that there should be perfect agreement. 
Figure \ref{fig:power-spectrum} shows the power spectrum obtained from each dataset individually and was calculated with the method described in \cite{IceCube:2012feb} and \cite{HAWC:2014dec}.
The missing $C_\ell$ for low $\ell$ is an artifact of the method that rises due to the limited sky coverage of the data.

%

\section{Discussion and Future Plans}
Current statistics for HAWC-250 are much lower than those of IC86 and do not comprise a full year. 
This limited coverage results in contamination from the solar dipole that can be observed in anti-sidereal coordinates.
While HAWC-250 has a higher trigger rate than IceCube,
restricting datasets to compatible energy bins reduces statistics for HAWC by nearly a factor of 10. 
This means that ${\sim}3$ years of HAWC-250 are needed in order to have statistics equivalent to 1 year of IC86 in the energy range of interest.

Additional work is needed in order to understand differences in the two datasets in terms of mass composition and energy distribution.
For example, the IceCube detector observes cosmic-ray air showers through
TeV muons which travel on the order of 1 km through the South Pole ice sheet. At the energy threshold of this 
analysis, the IceCube detector will preferentially trigger on
proton events over heavier nuclei (which produce lower energy
muons on average)~\cite{IceCube:2012feb}. Recent measurements also indicate that
the TeV band is a complex region of changing cosmic-ray
composition, with protons becoming the sub-dominant
primary type at energies above 10 TeV~\cite{hsahn2010,Adriani01042011}. 
As a result, IceCube and HAWC may not be observing a completely
equivalent population of cosmic rays. 

\section{Conclusions}

We have performed a preliminary analysis of the cosmic-ray anisotropy at TeV energies by 
combining datasets from the IceCube Neutrino Observatory, in the Southern Hemisphere, and the 
HAWC gamma-ray observatory, in the Northern Hemisphere. The objective of this ongoing study is to eliminate systematic effects and statistical uncertainties 
of the calculated angular power spectrum that result from partial sky coverage in each individual  dataset.

Using limited statistics from the 250-tank configuration of HAWC and a full year of the 86-string configuration of IceCube, 
we observe both significant large-scale and small-scale anisotropy in the arrival direction distribution of TeV cosmic rays but note that the HAWC data are contaminated by the
solar dipole that results from partial year coverage. 
It is expected that this main systematic bias should average out as we accumulate statistics and have a full year of HAWC data.
Work is underway to identify and eliminate additional sources of systematic biases and incompatibilities between the data from the two detectors.



\bibliographystyle{JHEP}
\bibliography{IC86_HAWC_CR_Aniso}


\end{document}